%% file: main.tex
\documentclass[a4paper, 11pt]{article}
\input{bib_macros}

\usepackage{geometry}
\usepackage[numbers]{natbib}
\usepackage[utf8]{inputenc}
\usepackage{enumitem,amssymb}
\usepackage{ragged2e}
\usepackage{setspace}
\usepackage{eso-pic,graphicx}
\usepackage{tikz}
\usepackage{caption}
\usepackage{lmodern}

\usepackage{xcolor}
\definecolor{cobalt}{rgb}{0., 0.35, 0.56}
\usepackage{hyperref}
\hypersetup{colorlinks=true,
            citecolor=cobalt,
            linkcolor=cobalt,
            urlcolor=cobalt,
            pdftitle={Standard Sirens in 2040s}}

\AddToShipoutPictureBG{%
  \ifnum\value{page}=1
    \AtPageLowerLeft{%
      \includegraphics[width=\paperwidth]{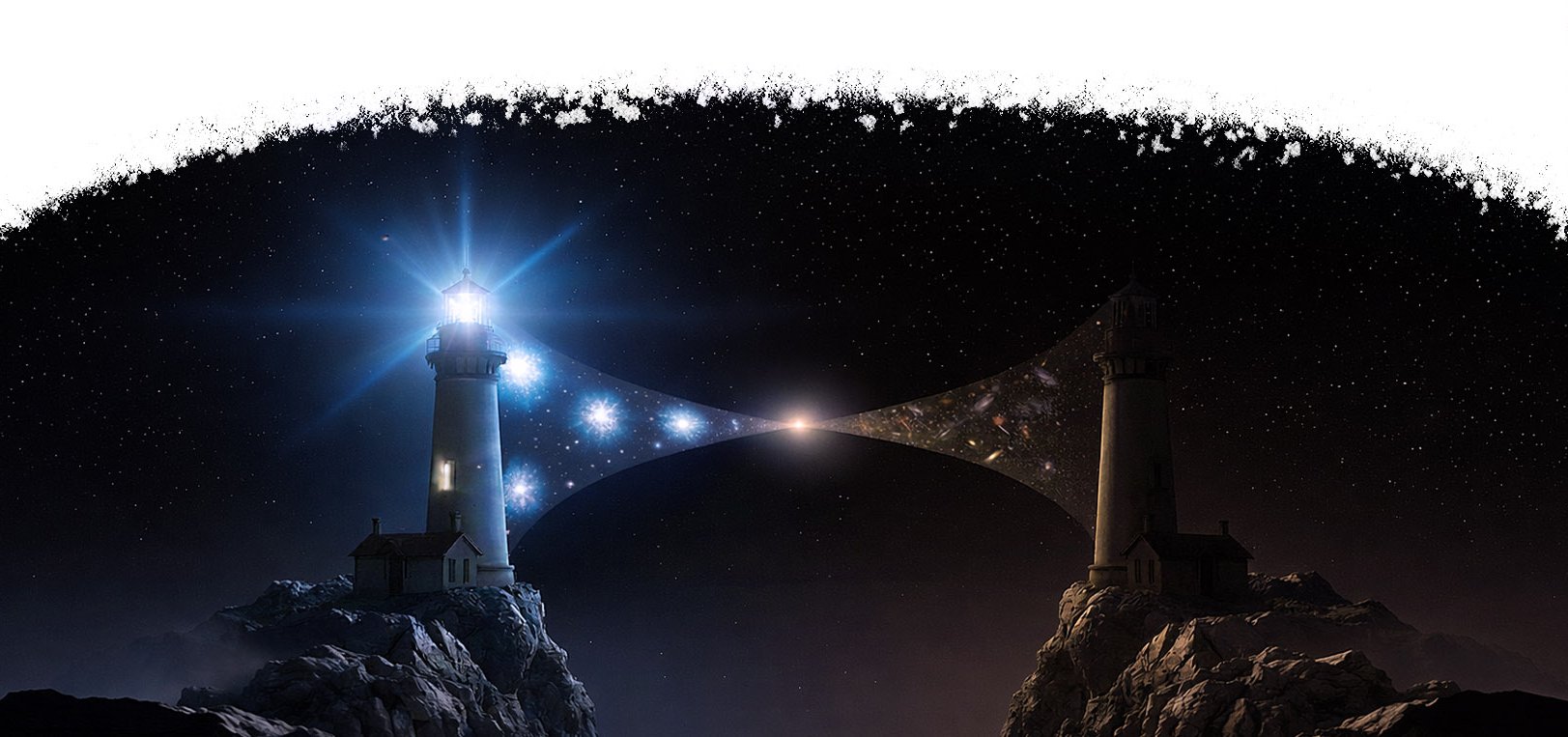}%
    }%
  \fi
}

\makeatletter
\patchcmd{\@lbibitem}
  {\item[\hfil\NAT@anchor{#2}{\NAT@num}]}
  {\item[\NAT@anchor{#2}{\textbf{\NAT@num}}]}
  {}{}
\makeatother

\begin{document}

\raggedright
\Large
ESO Expanding Horizons initiative \linebreak
Call for White Papers

\vspace{2.cm}
\begin{spacing}{1.6}
\textbf{\fontsize{22pt}{40pt}\selectfont
Standard Sirens in 2040s: Probing the Cosmic Expansion History with Gravitational Waves and Spectroscopic Galaxy Surveys}
\end{spacing}
\normalsize

\vspace{0.5cm}

\textbf{Authors:} Nicola Borghi$^\mathbf{1,2,3}$, Michele Moresco$^\mathbf{1,2}$, Richard I. Anderson$^\mathbf{4}$, Carmelita Carbone$^\mathbf{5}$, Andrea Cimatti$^\mathbf{1}$, Stephanie Escoffier$^\mathbf{6}$, Carlo Giocoli$^\mathbf{2,3}$, Sean MacBride$^\mathbf{7}$, Fatemeh Zahra Majidi$^\mathbf{8}$, Dinko Milakovi{\'c}$^\mathbf{9,10}$, Lauro Moscardini$^\mathbf{1}$, Lucia Pozzetti$^\mathbf{2}$, Margherita Talia$^\mathbf{1,2}$, Elena Tomasetti$^\mathbf{1,2}$

\vspace{0.3cm}

\textbf{Contact:} \href{mailto:nicola.borghi6@unibo.it}{nicola.borghi6@unibo.it}
\linebreak

\textbf{Affiliations:} \\
{\small 
$^\mathbf{1}$ Dipartimento di Fisica e Astronomia ``Augusto Righi'' -- Universit\`{a} di Bologna, via Gobetti 93/2, I-40129 Bologna, Italy \\
$^\mathbf{2}$ INAF-OAS, Osservatorio di Astrofisica e Scienza dello Spazio di Bologna, via Gobetti 93/3, I-40129 Bologna, Italy \\
$^\mathbf{3}$ INFN-Sezione di Bologna, Viale Berti Pichat 6/2, 40127 Bologna, Italy \\
$^\mathbf{4}$ Institute of Physics, \'Ecole Polytechnique F\'ed\'erale de Lausanne (EPFL), Observatoire de Sauverny, 1290 Versoix, Switzerland \\
$^\mathbf{5}$ INAF - Istituto di Fisica Spaziale e Fisica cosmica, Via Corti 12, I-20133 Milano (MI), Italy \\
$^\mathbf{6}$ Aix Marseille Universit\'e, CNRS/IN2P3, CPPM, Marseille, France \\
$^\mathbf{7}$ Physik-Institut, University of Zurich, Winterthurerstrasse 190, 8057 Zurich, Switzerland \\
$^\mathbf{8}$ INAF Capodimonte Astronomical Observatory, Salita Moiariello, 16, 80131 Naples, Italy \\
$^\mathbf{9}$ INAF-OATs, Osservatorio di Trieste, Via Tiepolo 11, I-34131 Trieste, Italy \\
$^\mathbf{10}$ Institute for Fundamental Physics of the Universe (IFPU), via Beirut 2, I-34151 Trieste, Italy 
}
\pagenumbering{gobble}

\pagebreak 

\justifying

\section*{Introduction}

Gravitational waves (GWs) from compact binary coalescences have matured into a robust cosmological probe, providing self-calibrated luminosity distance measurements independent of any cosmic distance ladder, hence the term ``standard sirens'' \citep{Schutz:1986gp}. The binary neutron star merger GW170817 delivered the first such measurement of the Hubble constant \citep{LIGOScientific:2017adf}, demonstrating that GWs offer a path to precision cosmology with systematics orthogonal to standard cosmological probes. 
To convert GW distances into cosmological parameters, redshift information is essential. To maximize the scientific potential, the redshift must be obtained from individual galaxies, either by identifying electromagnetic counterparts of GW events (bright sirens) or by statistically associating potential hosts within the GW localization volume (dark sirens). The precision of these redshifts sets the achievable accuracy. Forecasts show that photometric uncertainties degrade cosmological constraints by up to an order of magnitude compared to spectroscopic ones \citep{Borghi:2023opd, Tagliazucchi:2025ofb}. Wide-field, high-multiplex spectroscopic facilities will therefore be an essential infrastructure for GW cosmology in the 2040s.

\section*{Standard Siren Cosmology in the 2040s}

Ten years after the first detection, the LIGO-Virgo-KAGRA (LVK) collaboration has completed its fourth observing run (O4), with more than 200 candidate events publicly released, the vast majority being binary black holes (BBHs) \citep{LIGOScientific:2025slb}. In parallel, a significant effort of the community has been devoted to developing robust statistical methodologies to extract cosmological constraints from the growing GW catalog \citep{LIGOScientific:2025jau}.
With LVK progressing toward design sensitivity (O5, planned for the early 2030s) and third-generation detectors such as the Einstein Telescope (ET) and Cosmic Explorer (CE) planned for the 2040s, the observational landscape will transform. These facilities are expected to detect $10^4$--$10^5$ compact binary mergers per year at redshifts $z>2$ for BBHs and $z>3$ for binary neutron stars (BNSs) \citep{Branchesi:2023mws, ET:2025xjr}. The space-based LISA mission will open a complementary low-frequency window, targeting massive black hole binaries that can also be exploited for cosmology \citep{LISACosmologyWorkingGroup:2022jok}.

\noindent\textbf{Bright sirens} require electromagnetic counterparts such as kilonovae or gamma-ray burst afterglows to identify the host galaxy and measure its redshift directly. Such events are rare, as kilonovae reach peak $i$-band magnitudes $m_i\sim22-25$ at the distances probed by third-generation detectors and fade within days, while afterglow detection requires a favorable viewing geometry. With ET+CE, $\mathcal{O}(10)-\mathcal{O}(100)$ bright sirens per year could have well-localized counterparts \citep{Bisero:2025tkw}.

\begin{figure}[ht]
\centering
\includegraphics[width=\linewidth]{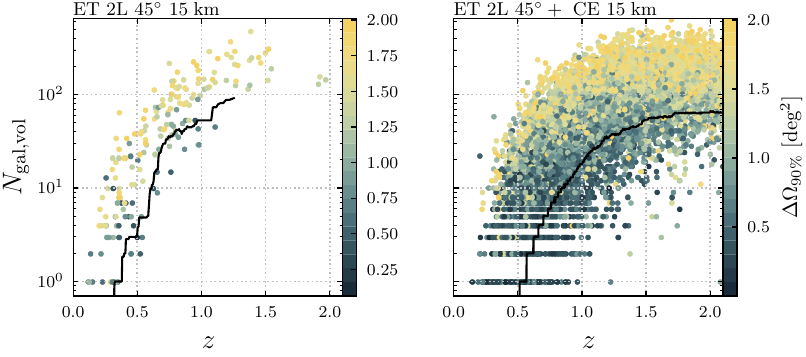}
\caption{Number of massive galaxies inside the localization volume of golden BBH events ($\Delta\Omega_{90\%} <2~\mathrm{deg^2}$) for ET (\textit{left}, 2L 15~km configuration) and ET+CE (\textit{right}, 15~km configuration). The galaxy catalog contains $\log M_\star/\mathrm{M_\odot}>10$ galaxies from the \textit{Euclid} Flagship simulations \citep{Euclid:2024few}. To compute $N_\mathrm{gal,vol}$ we adopt wide priors on $H_0\in[60,80]~\mathrm{km\,s^{-1}\,Mpc^{-1}}$. The black line and shaded region indicate the 84th percentile.}
\label{fig:Ngal}
\captionsetup{labelformat=empty}
\end{figure}

\noindent\textbf{Dark sirens}, the dominant BBH population without counterparts, rely on statistical association with galaxy catalogs. The method constructs a redshift prior from galaxies within the three-dimensional GW localization volume, weighted by merger probability. The number of galaxies per localization volume (Figure~\ref{fig:Ngal}) is a key metric. When only one galaxy lies within the error volume, a dark siren achieves bright-siren precision.

\noindent The science case for third-generation GW detectors spans multiple frontiers requiring deep and robust spectroscopic galaxy information:

\begin{itemize}
    \item \textit{Precision cosmology with $H_0$.} Current standard siren constraints on $H_0$ are at the $\sim$15\% level \citep{LIGOScientific:2025jau}. ET will localize $\mathcal{O}(10^2)$ BBH sources annually to better than $2\,\rm deg^2$ \citep{Branchesi:2023mws}, and for nearby, well-localized events the volume may contain a single massive galaxy, enabling sub-percent $H_0$ from individual dark sirens \citep{ET:2025xjr}. Moreover, bright sirens with kilonovae can enable sub-percent measurements in one year of ET observations \citep{Branchesi:2023mws}. This level of accuracy would be highly informative of the origin and implications of the $\sim 7\sigma$ discord between the global value of the Hubble constant inferred from the Cosmic Microwave Background assuming $\Lambda$CDM \citep{SPT-3G:2025bzu} and the directly measured Hubble constant based on the Local Distance Network \citep{H0DN:2025lyy}.
        
    \item \textit{Expansion history.} Mapping $H(z)$ beyond $z\sim 0.5$ requires the large event rates and precise localizations of third-generation networks. With ET+CE, well-localized dark sirens will extend to $z\sim 1-2$, probing the expansion history at earlier cosmic epochs. Sub-percent precision on $H(z)$ over more than 10 billion years of cosmic evolution becomes achievable when combining future bright sirens and golden dark sirens \citep{Chen:2024gdn}.
    
    \item \textit{Modified GW propagation.} Theories beyond general relativity (GR) can predict a friction term $\Xi(z)$ modifying the GW luminosity distance \citep{Belgacem:2018lbp}. Current constraints are promising ($\Xi_0 = 1.2^{+0.8}_{-0.4}$ from GWTC-4.0 data \citep{LIGOScientific:2025jau}). Since modified propagation is a cumulative effect increasing with distance, high-redshift sources accessible to ET provide the sensitivity needed to probe this effect with unprecedented precision. In particular, ET can reveal deviations from GR of order $10^{-2}$ \citep{ET:2025xjr}.
    
    \item \textit{Host environments and astrophysics.} Correlating GW source properties with host galaxy properties (e.g., stellar mass, star formation rate, metallicity) can be used to inform binary formation channels. At the same time, knowledge of host galaxy properties can reduce the effective number of candidate hosts in dark siren analyses, tightening cosmological constraints and reducing systematics.
\end{itemize}

\noindent\textbf{The critical role of spectroscopic redshifts.} Forecasts using hierarchical Bayesian inference demonstrate that with spectroscopic redshifts ($\sigma_z\sim0.001(1+z)$), 100 well-localized BBH events at LVK O5-like sensitivity constrain $H_0$ to \%-level precision. With photometric redshifts ($\sigma_z\sim0.05(1+z)$), the same sample achieves only 9\%, a factor of $\sim$10 degradation \citep{Borghi:2023opd}. For modified gravity, photometric uncertainties degrade constraints on $\Xi_0$ up to a factor of $\sim$5 \citep{Tagliazucchi:2025ofb}.

\section*{What Stage IV Cannot Deliver}

Current and planned Stage~IV spectroscopic surveys (e.g., DESI, 4MOST, MOONS) face fundamental limitations:

\begin{itemize}\itemsep0pt
    \item \textbf{Redshift coverage:} DESI reaches $z\sim 1.6$ for emission-line galaxies; 4MOST and MOONS target $z\lesssim1.5$. Beyond $z\sim2$, only sparse quasar samples exist. Yet ET will detect BBHs across this full range.
    
    \item \textbf{Depth and area:} Complete spectroscopic host galaxy catalogs to $m_i\sim25$ across $\gtrsim10{,}000\,\rm deg^2$ exceed Stage~IV capabilities. Current multi-object spectrographs lack the sensitivity and multiplex to survey the required galaxy populations efficiently.
    
    \item \textbf{Target-of-opportunity response:} Electromagnetic counterpart identification demands rapid spectroscopic follow-up ($<$24 hours) across GW localization volume containing up to $\mathcal{O}(10^3)-\mathcal{O}(10^4)$ candidates \citep{Bisero:2025tkw}. Existing facilities cannot deliver this performance at the required depths.
\end{itemize}

\section*{Requirements and Facility Concept}

Enabling the cosmological potential of third-generation GW detectors {\bf requires} a spectroscopic infrastructure with:

\begin{itemize}\itemsep0pt

    \item \textbf{Wide-area, deep spectroscopy:} Complete catalogs to $z\lesssim0.5$ at depths $m_i\sim25$, covering $\gtrsim10{,}000\,\rm deg^2$. This redshift range captures the majority of well-localized BBH events and most bright sirens. The depth requirement matches photometric surveys (Vera Rubin Observatory) that will identify transient candidates. Wide sky coverage ensures a more complete mapping of the observed GW events.
        
    \item \textbf{High multiplex:} Simultaneous spectroscopy of $\mathcal{O}(10^4)$ targets per pointing to efficiently survey dense galaxy fields within GW localization volumes. In the best-case scenario, a single well-localized BBH event at $z\sim 0.9$ contains $\mathcal{O}(10^2)$ massive galaxies (Figure~\ref{fig:Ngal}). Conventional facilities would require prohibitive observing time to achieve the required completeness for many events; high multiplex enables complete spectroscopic coverage of error volumes in single or few pointings.
    
    \item \textbf{Rapid response:} Target-of-opportunity capability for electromagnetic counterpart identification, combining integral-field and multi-object modes. Multi-object spectroscopy (MOS) allows simultaneous observation of hundreds to thousands of candidate hosts within GW error regions. Current facilities (4MOST, WEAVE) lack the combination of field-of-view, multiplex ($\sim$30{,}000 targets), sensitivity to $m_i\sim25$, and rapid ToO scheduling needed for this science.
    
    \item \textbf{Spectroscopic precision:} Redshift accuracy $\sigma_z/(1+z)\lesssim10^{-3}$ is required to break cosmology--population degeneracies. Spectroscopic precision yields a $\sim 10 \times$ improvement in $H(z)$ constraints compared to photometric redshifts for the same event sample. Beyond redshift measurements, spectroscopic surveys provide information on host galaxy properties (e.g., stellar mass, star formation rate, and metallicity) that can be incorporated as astrophysically motivated priors to improve cosmological inference.
     
\end{itemize}

\noindent The Wide-field Spectroscopic Telescope (WST), currently under concept study, exemplifies the required capabilities: a large collecting area, $\sim$3\,deg$^2$ field of view, and $\sim$20{,}000--30{,}000 fiber multiplex optimized for both survey and transient science \citep{WST:2024rai}. The scientific return from third generation GW observatories could be fundamentally limited by the quality of available redshift information. Spectroscopic surveys designed with GW cosmology as a core science driver constitute a critical infrastructure for precision cosmology in the 2040s.

\vfill 
% \small

\bibliographystyle{aa_1auth}
\bibliography{references}

\end{document}

%% file: bib_macros.tex
\usepackage{paralist}